\documentclass[aps,two column,showpacs,preprintnumbers,amsmath,amssymb]{revtex4}
\usepackage{graphicx}
\usepackage{subfigure}
\usepackage{dcolumn}
\usepackage{latexsym}

\usepackage[T1]{fontenc}
\usepackage{bm}

\begin{document}

\title{Generation of modal and path entangled photons \\using domain engineered integrated optical waveguide device}
\author{Jasleen Lugani}\email{jaslphy@gmail.com}
\affiliation{Department of Physics, IIT Delhi, New Delhi 110016, India}
\author{Sankalpa Ghosh}
\affiliation{Department of Physics, IIT Delhi, New Delhi 110016, India}
\author{K. Thyagarajan}
\affiliation{Department of Physics, IIT Delhi, New Delhi 110016, India}
\date{\today}
\begin{abstract}
Integrated optical devices are expected to play a promising  role in the field of quantum information science and technology. In this paper, we propose a scheme for the generation of  non-degenerate, co-polarized, modal and path entangled photons using a directional coupler and an asymmetric Y-coupler geometry  in type 0 phase-matched, domain engineered  lithium niobate (LN) waveguide. The nonlinearity in LN is tailored in such a way that quasi phase matching conditions for two different spontaneous parametric down conversion (SPDC) processes are obeyed simultaneously, leading to a modal and path-entangled state at the output. Assuming typical values of various parameters, we show, through numerical simulations, that an almost maximally entangled state is achievable over a wide range of waveguide parameters.  For the degenerate case, the proposed scheme gives a NOON state for \textit{N} = 2. The generated entangled photon pairs should have potential applications in quantum information schemes and also in quantum metrology. By appropriate domain engineering and component designing, the idea can be further extended to generate a hyperentangled and two-photon multi-path entangled states which may have further applications in quantum computation protocols.

\end{abstract}
\maketitle
\section{Introduction}
\label{intro}
Quantum entangled photons, owing to their unique properties, play a crucial role in the implementation of quantum information protocols, quantum teleportation~\cite{Bouwmeester, Kimble, Furusawa}, and quantum computing~\cite{Walther1, EKnill}. The entanglement of photons can be exhibited in terms of their polarization which is intrinsically binary, or spatial and spectral degrees of freedom which are continuous variables. An important class of entangled photon states is NOON states or path-entangled states, defined as:$\left|\mathcal{N}\right\rangle=\frac{1}{\sqrt{2}}\left(\left|N,0\right\rangle_{a,b}+\left|0,N\right\rangle_{a,b}\right)$which is the coherent  superposition of  \textit{N} photons in two different spatial modes \textit{a} and \textit{b}. Nowadays, such NOON states are being exploited to beat the classical Rayleigh diffraction limit and thus, believed to find potential applications in quantum metrology, imaging and lithographic techniques~\cite{Dowling, PKok, Angelo}.

  Spontaneous parametric down conversion (SPDC) process in second order $\left(\chi^{2}\right)$ nonlinear bulk crystals and waveguides is a well established method for the generation of polarization entangled ~\cite{Kwait,Fiorentino,Fedrizzi, Martin} and time-bin entangled photons~\cite{Tanzilli,Honjo}. The use of waveguide structures leads to enhanced nonlinear efficiencies due to tight confinement of the interacting waves and in addition, provides a means to control the spatial characteristics of the down converted photons by confining them to well defined discrete transverse spatial modes~\cite{Ren,Eckstein}. This is in contrast to the continuous spatial distribution governed by momentum conservation in bulk crystals. By restricting the modes of the waveguide structure to the two lowest order modes, the mode number can serve as a modal qubit~\cite{Saleh1}, which can be used as an alternative to polarization qubit. The guided-wave downconverted photon pairs can thus be emitted in the two allowed modes leading to modal entanglement. Such waveguide structures are also compatible with integrated optics technology and form basic components in photonic circuit designs, which are expected to play an important role in quantum information technology, owing to the compactness of the device and low loss. The generation, separation and processing of entangled photons  can be carried out on a single chip and hence, the entangled photon state is less susceptible to decoherence. The photonic circuits based on titanium indiffused channel LN waveguides have also been recently investigated for different combinations of modal, spectral and polarization entangled photon pairs~\cite{Saleh2, Saleh3}.  
  
   In this paper, we address the issue of generation of a modal-path entangled state using SPDC in a domain engineered integrated optical waveguide device. Several schemes have been proposed over the years to generate two photon and multiphoton path entanglement using SPDC process with the use of interferometric set ups consisting of polarization beam splitters (PBS) or Mach Zehnder interferometers (MZI)~\cite{Gerry, Walther2, Sauge}, cavity QED~\cite{Dowling2}. We propose a scheme for  the generation of non-degenerate, co-polarized modal entangled state using a directional coupler in type 0 phase-matched doubly periodically polarized lithium niobate (LN) waveguide. The nonlinearity in LN is tailored in such a way that quasi phase matching (QPM) conditions for two different SPDC processes are obeyed simultaneously ~\cite{KT}. These two SPDC processes, when enabled together, lead to a mode entangled output state. We have shown through quantum mechanical analysis and numerical simulations, that for our choice of waveguide geometry and the modes, maximally mode entangled photon pairs can be achieved for a wide range of input parameters of the waveguide. Using an asymmetric Y-coupler at the output, it is then possible to convert the mode entangled state to a path entangled state. The use of the directional coupler  in our scheme  facilitates the choice of the modes required, which leads to an appreciable overlap of the fields at the pump, signal and idler wavelengths. The overlap integrals, which primarily determine the down conversion efficiency, are quite high in comparison to a two mode waveguide, giving rise to a high photon pair generation rate. Also, due to the design of the structure, we show that an efficient generation of photon pairs is possible using the fundamental spatial mode of the pump as compared to the earlier work in which the first excited spatial mode had to be chosen in order to achieve good efficiencies ~\cite{Saleh1}. The proposed device is flexible and the waveguide parameters can be modified to generate a hyperentangled state or a multi-path entangled state. 
      
       The organization of the paper is as follows. In Sec.~\ref{wavegeo}, we describe the theory and the design of the integrated optical waveguide geometry considered for the generation of mode-entangled state and the conversion of it to a path-entanglement. In Sec.~\ref{analysis}, we give a brief account of the basic quantum mechanical analysis for the two coupled SPDC processes involved in the generation of the desired mode-entangled state. In Sec.~\ref{simulations}, we present the results of numerical simulations and quantify entanglement by calculating the von Neumann entropy of the output state and show that a maximally entangled state is achievable using the proposed device. We also investigate the emission bandwidths of the two corresponding SPDC processes. We conclude with a brief summary in Sec.~\ref{conclusion}.  
\section{Waveguide Geometry}
\label{wavegeo}
\begin{figure}[h!]
\includegraphics [scale=0.355]{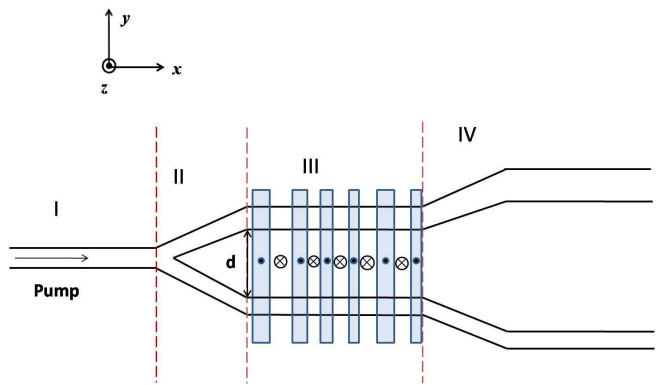}
\caption{(Color online) Waveguide design for generating modal and path-entangled photons. Region I is a single mode waveguide at pump wavelength, region II is a symmetric Y-splitter, region III is a directional coupler consisting of doubly periodic grating in lithium niobate (LN) substrate (dots and crosses represent the polarizing directions in the grating). Region IV is an asymmetric Y coupler. Pump, signal and idler are all \textit{z}-polarized, \textit{z} axis being the optic axis for the LN substrate.}
\label{Fig_wavegeom}
\end{figure}

We seek to generate a mode entangled state using spontaneous parametric down conversion process in which the signal and idler photons get generated in an entangled pair of modes. Once generated, such a mode- entangled state can be converted to a path-entangled state, such that, if one of the down converted photons comes out of one of the two (or more) available paths, the other photon (generated with the first photon) would also come out of the same path and similarly for the other path(s). In order to achieve this,  we consider a waveguide geometry as shown in Fig.~\ref{Fig_wavegeom}. Region I is a single mode waveguide at the pump wavelength, followed by region II consisting of a symmetric Y-splitter. Region III is a directional coupler supporting the fundamental (symmetric) normal mode and the first excited (antisymmetric) normal mode at signal and idler wavelengths. Using an appropriate phase matching in the directional coupler region, it is possible to generate down converted photons which are entangled in mode number.
  
    The working of the proposed device is as follows: When the fundamental mode at pump enters the input waveguide, it passes through the symmetric Y-splitter and excites the fundamental symmetric normal mode at pump wavelength in the directional coupler (region III). 
  The directional coupler region is assumed to have a doubly periodic domain reversal grating which simultaneously satisfies the quasi phase matching (QPM) conditions~\cite{KT} for the following two different SPDC processes: pump photon in the fundamental normal spatial mode down converting either a) both in the fundamental normal spatial mode or b) both in the first excited normal spatial mode. The substrate used in this region is assumed to be a domain engineered LN substrate which results in type 0 phase matched down conversion. 
     
     We show in Sec.~\ref{analysis}, that when the QPM conditions for both the  SPDC processes are satisfied simultaneously, it will lead to the generation of a  mode entangled signal-idler  photon pair in the doubly reversed region. Thus, in contrast to the standard phase matching based on a single process either in the bulk or in the waveguide configuration, domain engineering  can lead to the direct production of co-polarized mode-entangled state without the need for an interferometric set up or a 50:50 beam splitter .
    
    To convert the mode-entangled state to a path-entangled state, at the output of the domain reversed  region, we have assumed an asymmetric Y-coupler in region IV with both the output waveguides being singlemoded and the propagation constant of the upper waveguide being higher than the lower waveguide for signal and idler frequencies. Thus the fundamental normal modes of signal and idler exit from the upper waveguide while the first excited normal modes exit from the lower waveguide leading to a path-entangled state. We show through numerical simulations that with an appropriate choice of parameters of the coupler, it is possible to optimize the generation of the entangled state. In Sec.~\ref{analysis}, we present a brief account of the quantum mechanical analysis leading to the generation of two-photon two-mode entangled state in the direction coupler region (region III of Fig. ~\ref{Fig_wavegeom}). 
\section{Quantum mechanical analysis}
\label{analysis}
     We consider the process of parametric down conversion in a \textit{z}-cut, \textit{x}-propagating, domain engineered lithium niobate waveguide. We assume the pump, signal and idler wavelengths to be all having extraordinary polarization to make use of the largest non-linear coefficient ($d_{33}$) of lithium niobate. For the considered pump powers, the pump field  is assumed to be a classical field and the signal and idler fields are represented by quantum mechanical operators. The electric field distributions for pump [fundamental normal mode(0)], signal[fundamental(0) and first excited(1) normal modes] and idler[fundamental(0) and first excited(1) normal modes) are given by\\ 
Pump(0):
\begin{equation} 
\vec{E}_{p0}=\frac{1}{2}e_{p0}\left(\vec{r}\right)E_{p0}(e^{i(\beta_{p0}x-\omega_{p}t)}+e^{-i(\beta_{p0}x-\omega_{p}t)})\hat{z}\label{Eq_01}
\end{equation} 
Signal(0):
\begin{equation}
\hat{E}_{s0}=i\int d \omega_{s} e_{s0}\left(\vec{r}\right)\sqrt{\frac{\hbar\omega_{s}}{2\epsilon_{s0}\textit{L}}}(\hat{a}_{s0}e^{i\beta_{s0}x}-\hat{a}^{\dagger}_{s0}e^{-i\beta_{s0}x})\hat{z}\label{Eq_02}
\end{equation}
Signal(1):
\begin{equation}
\hat{E}_{s1}=i\int d \omega_{s}e_{s1}\left(\vec{r}\right)\sqrt{\frac{\hbar\omega_{s}}{2\epsilon_{s1}\textit{L}}}(\hat{a}_{s1}e^{i\beta_{s1}x}-\hat{a}^{\dagger}_{s1}e^{-i\beta_{s1}x})\hat{z}\label{Eq_03}
\end{equation}
Idler(0):
\begin{equation}
\hat{E}_{i0}=i\int d \omega_{i}e_{i0}\left(\vec{r}\right)\sqrt{\frac{\hbar\omega_{i}}{2\epsilon_{i0}\textit{L}}}(\hat{a}_{i0}e^{i\beta_{i0}x}-\hat{a}^{\dagger}_{i0}e^{-i\beta_{i0}x})\hat{z}\label{Eq_04}
\end{equation}
Idler(1):
\begin{equation}
\hat{E}_{i1}=i\int d \omega_{i}e_{i1}\left(\vec{r}\right)\sqrt{\frac{\hbar\omega_{i}}{2\epsilon_{i1}\textit{L}}}(\hat{a}_{i1}e^{i\beta_{i1}x}-\hat{a}^{\dagger}_{i1}e^{-i\beta_{i1}x})\hat{z}\label{Eq_05}
\end{equation}
where the subscripts 0 and 1 refer to the fundamental and the first excited modes respectively, $e_{p0}\left(\vec{r}\right)$, $e_{s0}\left(\vec{r}\right)$ and $e_{i0}\left(\vec{r}\right)$ represent the transverse dependence of the fundamental modal fields of pump (\textit{p}), signal(\textit{s}) and idler(\textit{i}), respectively and $e_{s1}\left(\vec{r}\right)$ and $e_{i1}\left(\vec{r}\right)$ represent transverse dependence of the first excited modal fields for signal(\textit{s}) and idler (\textit{i}) wavelengths, \textit{L} represents the interaction length, $\epsilon_{s}$ and $\epsilon_{i}$ correspond to the optical dielectric permittivity of LN for signal and idler; $\hat{a}_{s0}$ ($\hat{a}_{s1}$) and  $\hat{a}_{i0}$ ($\hat{a}_{i1}$) represent the quantum mechanical annihilation operators corresponding to the fundamental mode (first excited mode) of signal and idler. $\beta_{\textit{q}0}$ (\textit{q} = \textit{p}, \textit{s}, \textit{i}) corresponds to the propagation constant for the fundamental mode and $\beta_{\textit{r}1}$ (\textit{r} = \textit{s}, \textit{i}) corresponds to the propagation constant for the first excited mode at the respective wavelengths. Energy is conserved for both the processes such that $
\omega_{p}=\omega_{s}+\omega_{i}$;  $\omega_{p}$, $\omega_{s}$ and $\omega_{i}$ being the pump, signal and idler frequencies, respectively.

In order that the output state is entangled, the nonlinearity of LN substrate in the directional coupler region is engineered to satisfy the following two quasi phase matching (QPM) conditions, simultaneously, 
\begin{equation}
K_{1}=\frac{2\pi}{\Lambda_{1}}=\beta_{p0}-\beta_{s0}-\beta_{i0}\label{Eq_07}
\end{equation}
\begin{equation}
K_{2}=\frac{2\pi}{\Lambda_{2}}=\beta_{p0}-\beta_{s1}-\beta_{i1}\label{Eq_08}
\end{equation}
where $\Lambda_{1}$ and $\Lambda_{2}$ are the QPM grating periods for the individual SPDC processes. The second order non linear polarization generated in the medium is given by 
\begin{equation}
P^{NL}_{i}=2\epsilon_{0}\sum_{j,k}d_{ijk}E_{j}E_{k}\label{Eq_09}
\end{equation}
The indices \textit{i}, \textit{j}, \textit{k} (1, 2, 3) corresponds to \textit{x}, \textit{y}, \textit{z} components of the co-ordinate system. Thus, $E_j$ represents the $j^{th}$ component of the total electric field within the medium and $d_{ijk}$ is the corresponding non-linear coefficient. For the case under consideration, the components of the total electric field are
\begin{equation}
\begin{array}{l}
E_{1} = 0,\\ 
E_{2} = 0,\\ 
E_{3} = \hat{E}_{p0}+\hat{E}_{s0}+\hat{E}_{s1}+\hat{E}_{i0}+\hat{E}_{i1},
\end{array}
\label{Eq_11}
\end{equation}
 The interaction Hamiltonian is derived as ~\cite{KT}
\begin{equation}
\hat{H}_{int}=-4\epsilon_{0}\int\int\int d_{33}(\hat{E}_{p0}\hat{E}_{s0}\hat{E}_{i0}+\hat{E}_{p0}\hat{E}_{s1}\hat{E}_{i1})dxdydz\label{Eq_12}
\end{equation}
 Using the expressions for the electric fields defined in Eq.~(\ref{Eq_01}-\ref{Eq_05}) and under rotating wave approximation (RWA) and energy conservation, we obtain the following expression for the interaction Hamiltonian: 
 \begin{widetext}
\begin{eqnarray}
\hat{H}_{int} = \int d \omega_{s} \left(\frac{E_{p0} \hbar \sqrt {\omega_{s} \omega_{i}}}{L}\right) \int^{L}_{0}d_{33} \left[ \left( \frac{I_{0}}{n_{s0}n_{i0}}\right) \left(\hat{a}^{\dagger}_{s0}\hat{a}^{\dagger}_{i0}e^{i((\beta_{p0}-\beta_{s0}-\beta_{i0})x-\omega_{p}t)} + \hat{a}_{s0}\hat{a}_{i0}e^{-i((\beta_{p0}-\beta_{s0}-\beta_{i0})x-\omega_{p}t)}\right)\right. \nonumber \\
\left.+\left (\frac{I_{1}}{n_{s1}n_{i1}} \right) \left(\hat{a}^{\dagger}_{s1}\hat{a}^{\dagger}_{i1} e^{i((\beta_{p0}-\beta_{s1}-\beta_{i1})x-\omega_{p}t)} + \hat{a}_{s1}\hat{a}_{i1}e^{-i((\beta_{p0}-\beta_{s1}-\beta_{i1})x-\omega_{p}t)}\right)\right] dx \label{Eq_13}
\end{eqnarray}
\end{widetext}
$I_{0}$ and $I_{1}$ denote the overlap integrals of the electric field profiles of the pump, signal and idler modes involved in the two SPDC processes and are given as
\begin{eqnarray}
I_{0}=\int\int e_{p0}\left(\vec{r}\right)e_{s0}\left(\vec{r}\right)e_{i0}\left(\vec{r}\right)dydz, \label{Eq_14}\\
I_{1}=\int\int e_{p0}\left(\vec{r}\right)e_{s1}\left(\vec{r}\right)e_{i1}\left(\vec{r}\right)dydz
\label{Eq_15}
\end{eqnarray}
The domain engineering in LN substrate results in two independent spatial frequency components in the non linear coefficient variation, along the propagation direction ~\cite{KT}. Thus, the effective non linear coefficient \textit{d} including the effect of periodic domain reversal is given as 
\begin{eqnarray}
\bar{d}_{33}=-\frac{4d_{33}}{\pi^{2}}(e^{iK_{1}x}+e^{-iK_{1}x}-e^{iK_{2}x}-e^{-iK_{2}x})\nonumber\\
+\mbox{terms at other spatial frequencies}
\label{Eq_16}
\end{eqnarray}  
Replacing $d_{33}$ by $\bar{d}_{33}$ in Eq.~(\ref{Eq_13}) and simplifying, the interaction Hamiltonian in the interaction picture is obtained as
\begin{equation}
\hat{H}_{int}=\int d\omega_{s}(C^{'}_{0}(\hat{a}^{\dagger}_{s0}\hat{a}^{\dagger}_{i0}+\hat{a}_{s0}\hat{a}_{i0})+C^{'}_{1}(\hat{a}^{\dagger}_{s1}\hat{a}^{\dagger}_{i1}+\hat{a}_{s1}\hat{a}_{i1}))
\label{Eq_17}
\end{equation}
\\where
\begin{eqnarray}
C^{'}_{0}=-\left(\frac{4d_{33}\hbar\sqrt{\omega_{s}\omega_{i}}I_{0}E_{p0}}{\pi^{2}n_{s0}n_{i0}}\right)e^{\frac{-i\Delta k_{0} L}{2}}\mbox{sinc}\left(\frac{\Delta k_{0} L}{2}\right)
\label{Eq_18}\\
C^{'}_{1}=-\left(\frac{4d_{33}\hbar\sqrt{\omega_{s}\omega_{i}}I_{1}E_{p0}}{\pi^{2}n_{s1}n_{i1}}\right)e^{\frac{-i\Delta k_{1} L}{2}}
\mbox{sinc}\left(\frac{\Delta k_{1} L}{2}\right)
\label{Eq_19}
\end{eqnarray}
\\and $\Delta k_{0}$ and $\Delta k_{1}$ are phase mismatch for the two processes given by 
\begin{eqnarray}
\Delta k_{0}=K_{1}-\beta_{p0}+\beta_{s0}+\beta_{i0}\\
\Delta k_{1}=K_{2}-\beta_{p0}+\beta_{s1}+\beta_{i1}
\label{Eq_21}
\end{eqnarray}
The output two photon state is found as
\begin{equation}
\left|\psi\right\rangle=\int d\omega_{s}i(C_{0}\left|s_{0},i_{0}\right\rangle+C_{1}\left|s_{1},i_{1}\right\rangle)
\label{Eq_22}
\end{equation}
and is entangled; $C_{0}=-\frac{t}{h}C^{'}_{0}$ and $C_{1}=-\frac{t}{h}C^{'}_{1}$.
The relative magnitudes of $C_{0}$ and $C_{1}$ coefficients will determine if the state is  maximally entangled or not. These, in turn, depend upon the overlap integrals and the effective indices of the interacting modes at pump, signal and idler wavelengths. In Sec.~\ref{simulations}, we carry out simulations for a planar waveguide and using practical values of waveguide parameters, we investigate various aspects of the output entangled state.
\section{Numerical Simulations}
\label{simulations}
\begin{figure}[h!]
\includegraphics[scale=0.4]{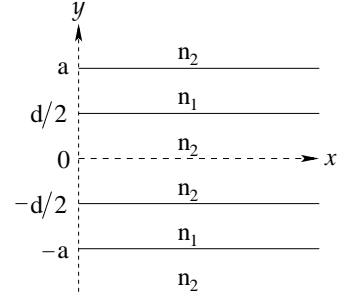}
\caption{Waveguide directional coupler geometry corresponding to Region III of Fig.~\ref{Fig_wavegeom}.}
\label{Fig_refindex}
\end{figure}

In this section, we report on the results of simulations on the entangled states generated in the directional coupler region using Titanium indiffused domain engineered planar lithium niobate waveguide. We model region III as shown in Fig.~\ref{Fig_refindex}. In order to demonstrate the generation of the entangled state in this region, we assume the planar waveguides to have step refractive index profiles. Since actual devices will use channel waveguides, our analysis can be used by first representing the channel waveguides by equivalent planar waveguides using the standard effective index method~\cite{Suhara}. The analysis presented can be easily extended to channel waveguide geometries with graded refractive index profiles; however, the general conclusions of our analysis would still be valid.
As shown in Fig.~\ref{Fig_refindex}, the guiding regions $(d/2 < \left|y\right| < a)$ are assumed to have refractive index  $n_{1}$ and the surrounding medium to have a refractive index $n_{2} (< n_{1})$. Thus, the geometry acts like a directional coupler with the cores of the individual waveguides, separated by distance \textit{d}. The individual core widths are $(a-d/2) = r$. Modal analysis for the coupler region was carried out ~\cite{Thyagarajan} and the eigenvalue equations for the symmetric and antisymmetric modes were obtained. The solution of these eigenvalue equations gives the propagation constants of all the modes, at all the wavelengths.
For the numerical simulations, the refractive indices of the substrate at the corresponding wavelengths have been calculated using temperature dependent Sellmeier equation for LN ~\cite{URL}. The pump, signal and idler wavelengths are chosen to be 750 nm, 1452 nm and 1550 nm, the corresponding refractive index differences $(n_{1}-n_{2})$ are assumed to be 0.0033, 0.0026 and 0.0025, respectively ~\cite{KT, Foutchet}. 
\begin{figure}[htbp]
\centering
\subfigure[]{
\includegraphics[scale=.235]{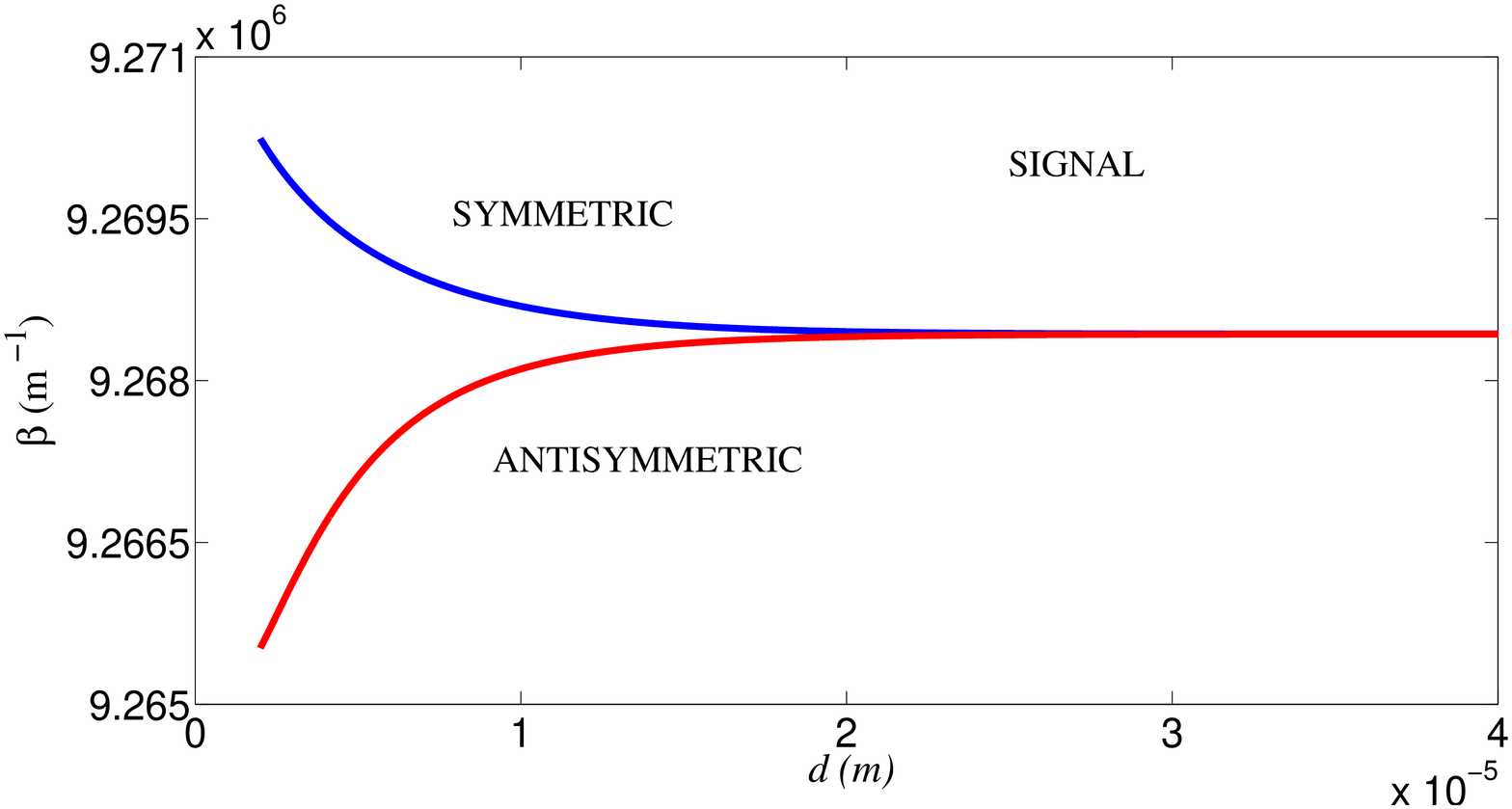}\label{fig:signal}}
\subfigure[]{
\includegraphics[scale=.24]{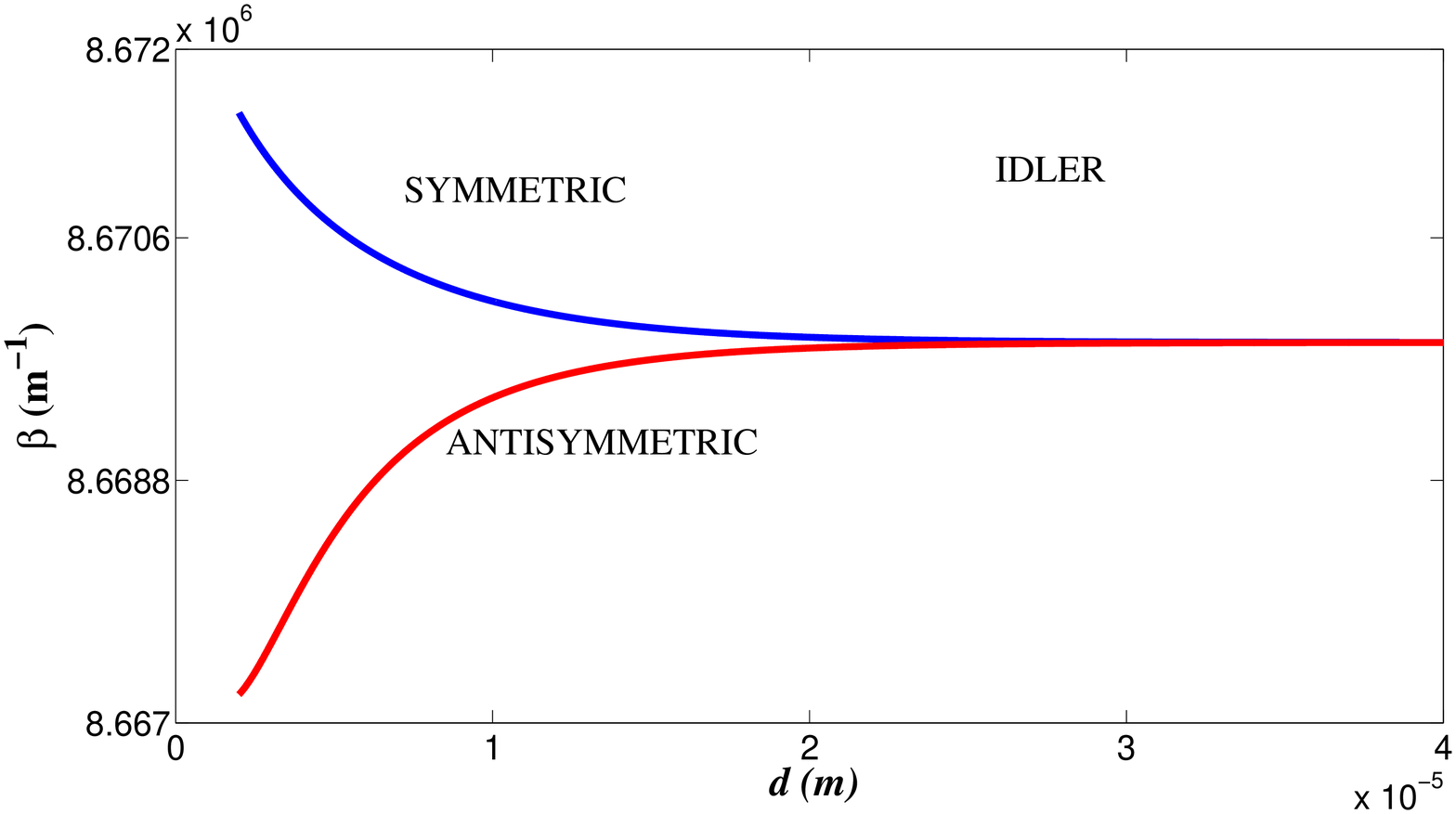}\label{fig:idler}}
\caption{(Color online) Variation of propagation constant with the core separation \textit{d} corresponding to (a) signal and (b) idler.}
\label{fig_propconstant}
\end{figure}
The propagation constants evaluated for the symmetric and antisymmetric modes at signal and idler wavelengths vary with the core separation \textit{d} as shown in Fig.~\ref{fig_propconstant}. We can see from Fig.~\ref{fig_propconstant} that, as expected, beyond a certain core  separation in the coupler region, the symmetric and antisymmetric modes have almost the same propagation constants for both signal and idler wavelengths.
\begin{figure}[h!]
\centering
\subfigure[]{\label{subfig_fundamental}\includegraphics[scale=.25]{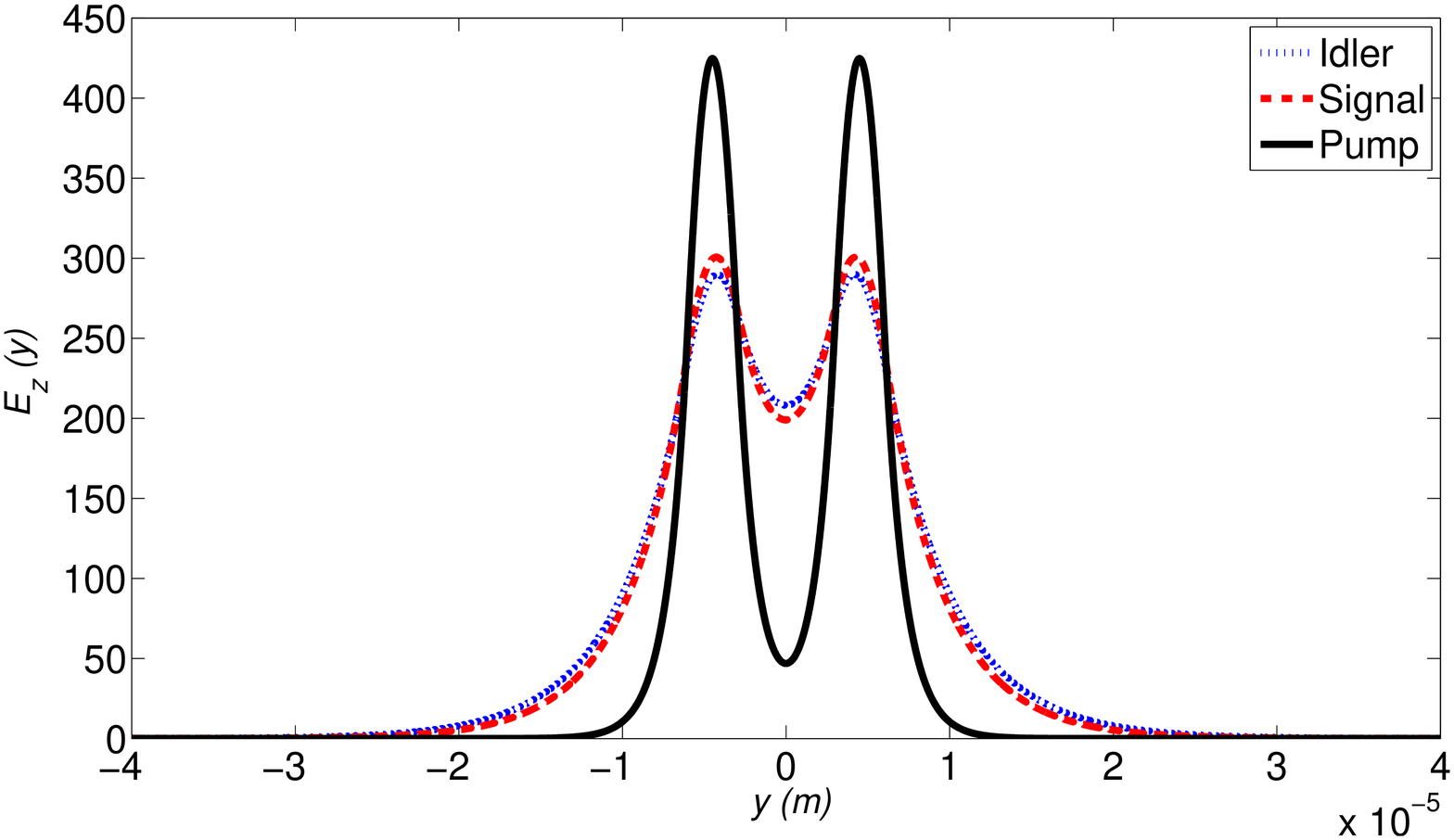}}
\subfigure[]{\label{subfig_first_excited}
\includegraphics[scale=.24]{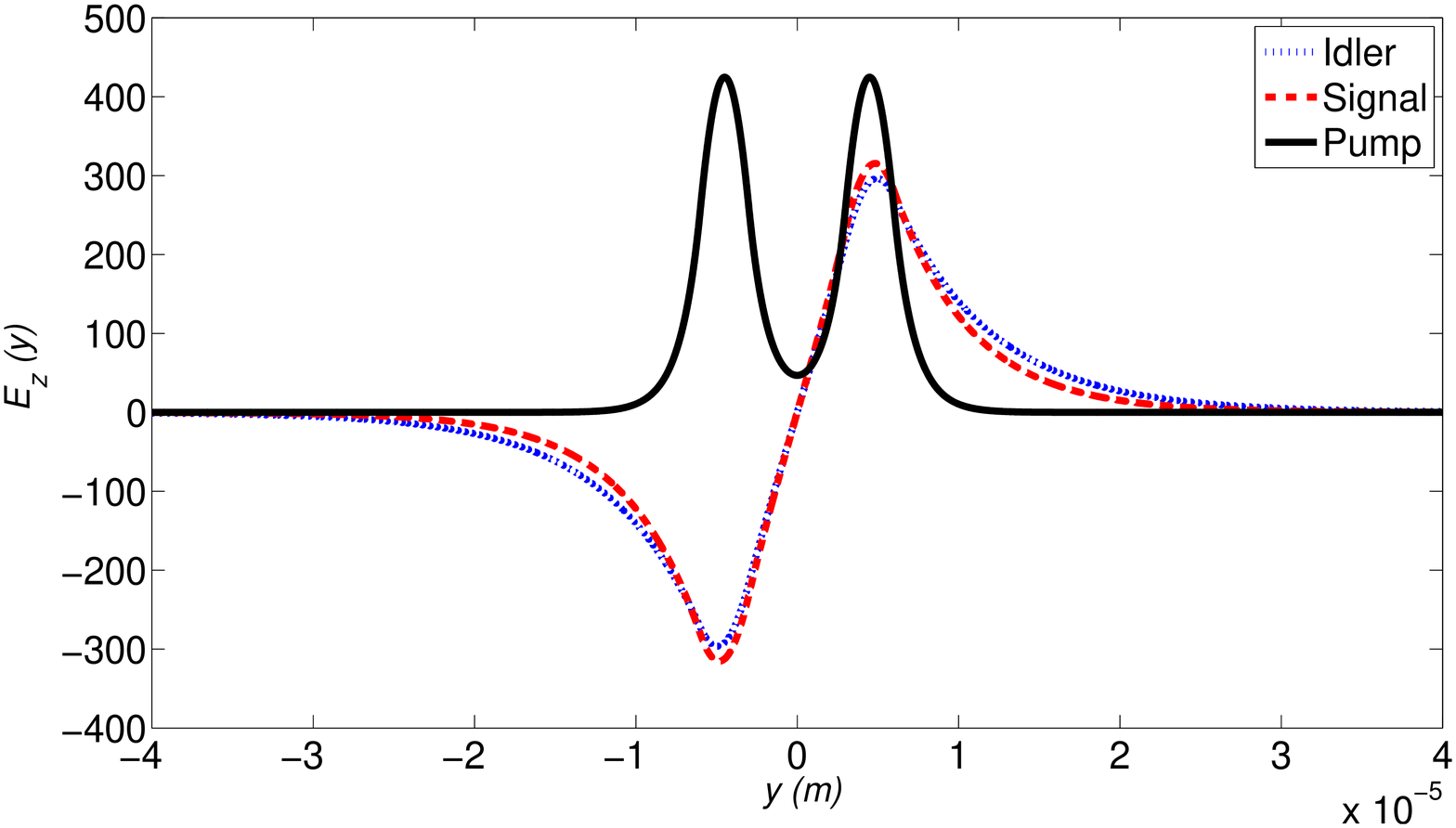}}
\caption{(Color online) Normalized transverse electric field distributions in the direction coupler region corresponding to (a) fundamental symmetric normal modes at pump, signal and idler wavelengths, and(b) fundamental symmetric normal mode at the pump and first excited antisymmetric normal mode at signal and idler wavelengths.}
\label{Fig_fieldprofiles}
\end{figure}     
            
            The transverse electric field distributions of the interacting modes at pump, signal and idler are shown in Fig.~\ref{Fig_fieldprofiles} for a core separation of 6 $\mu m$. Fig.~\ref{subfig_fundamental} shows the electric field distributions corresponding to the fundamental symmetric modes in the directional coupler region at the pump (solid), signal (dashed) and idler (dotted) wavelengths and  Fig.~\ref{subfig_first_excited} shows the field distributions corresponding to the fundamental symmetric mode of the pump (solid) and the first excited antisymmetric modes  at signal (dashed) and idler (dotted) wavelengths. The pump field is more tightly confined than the signal and idler wavelengths due to its shorter wavelength. Due to the choice of the waveguide geometry and wavelengths, it can be seen from the figure that there is a good overlap between the fields of the interacting modes at the respective wavelengths, resulting in high down conversion efficiencies for both the SPDC processes. 
\begin{figure}[h!] 
\includegraphics[scale=0.235]{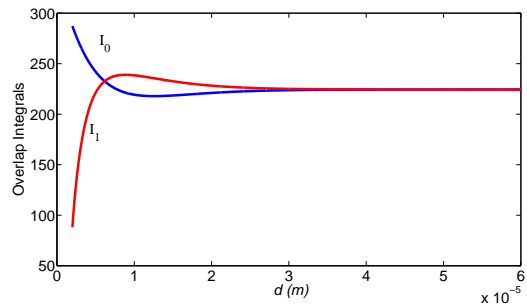}
\caption{(Color online) Variation of the overlap integrals with the core separation \textit{d}.}
\label{Fig_overlap}
\end{figure}
 From the field expressions for the spatial modes at pump, signal and idler, we have evaluated analytical expressions for the overlap integrals $I_{0}$ and $I_{1}$ (defined by Eq.~(\ref{Eq_14}) and Eq.~(\ref{Eq_15})) which are plotted with respect to the core separation in Fig.~\ref{Fig_overlap}. The figure shows that by an appropriate choice of the waveguide separation, it is possible to obtain the desired ratios of the overlap integrals that lead to a high degree of entanglement between the downconverted photons.
 
\begin{figure}[h]
\includegraphics[scale=0.275]{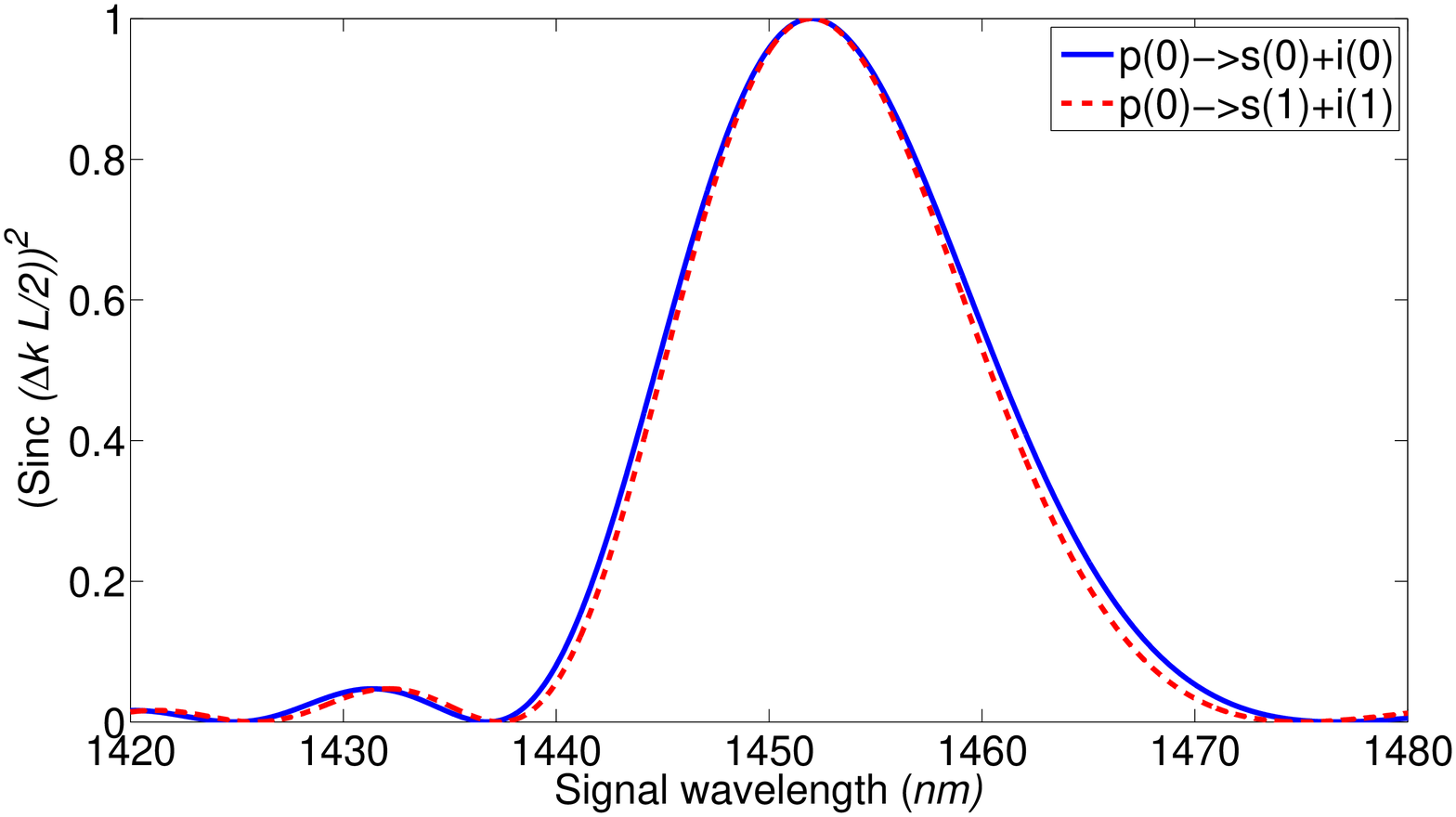}
\caption{(Color online) Normalized output signal spectra corresponding to the two SPDC processes considered.}  
\label{Fig_bandwidth}
\end{figure}

     In order to quantify the entanglement, we calculate Von Neumann entropy~\cite{Nielson} for the case of perfectly phase matched output entangled state of Eq.~(\ref{Eq_22}) (after normalization)  as $S = -tr(\rho^{s} \log_{2} \rho^{s})$, (tr denoting the trace and $\rho^{s}$ is the reduced density matrix of the signal (s) photon) and is derived as 
\begin{equation}
S=-\frac{I^{2}_{0}}{I^{2}_{0}+I^{2}_{1}}\log_{2}\frac{I^{2}_{0}}{I^{2}_{0}+I^{2}_{1}}-\frac{I^{2}_{1}}{I^{2}_{0}+I^{2}_{1}}\log_{2}\frac{I^{2}_{1}}{I^{2}_{0}+I^{2}_{1}}
\label{Eq_26}
\end{equation}
\begin{figure}[h!]
\includegraphics[scale=0.24]{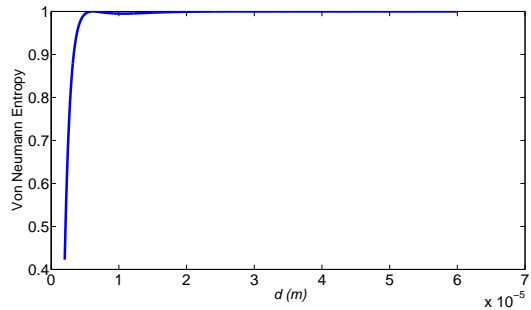}
\caption{(Color online)Variation of Von Neumann entropy with the core separation \textit{d}.}
\label{Fig_Von_Neumann}
\end{figure} 
 The variation of \textit{S} with the waveguide separation is plotted in Fig.~\ref{Fig_Von_Neumann}. Von Neumann entropy (\textit{S}) for a maximally entangled state is 1 and 0 for a product state. We can see from Fig.~\ref{Fig_Von_Neumann}, that for the entangled state generated using the proposed waveguide device, \textit{S} becomes almost unity beyond a waveguide separation of about 6 $\mu m$, thus the device is capable of producing a maximally entangled state over a wide range of waveguide separation. The corresponding grating periods required for the two SPDC processes at this core separation are 17.521  $\mu m$ and 17.358 $\mu m$ respectively, as calculated from Eq.~(\ref{Eq_07}) and Eq.~(\ref{Eq_08}). Another condition for obtaining a maximally entangled state with high efficiency requires identical bandwidths for the two enabled SPDC processes ~\cite{KT}. Fig.~\ref{Fig_bandwidth} shows the output spectra at signal wavelength corresponding to both the SPDC processes. It can be seen from the figure that the two processes have almost identical bandwidths (16 nm) ensuring a maximally entangled state over the region of overlap ~\cite{Saleh1}.

       We may mention here that the above simulations have been carried out using a titanium indiffused waveguide; simulations can also be performed for  proton exchange (PE) LN  waveguides which may further increase the efficiencies due to tighter confinement of modes.
   Thus in region III, we  achieve  two photon modal entanglement. In the proposed device, this mode entangled state further gets converted to a path entangled state, when the photons reach region IV, consisting of an asymmetric Y-coupler where owing to the difference in the  propagation constants of the upper and the lower waveguides, the fundamental modes of signal and idler would exit from the upper waveguide while the first excited modes would exit from the lower waveguide. This leads to a path entangled state of the form
\begin{equation}
\left|\psi\right\rangle=\int d\omega_{s}i(C_{0}\left|s_{u},i_{u}\right\rangle+C_{1}\left|s_{l},i_{l}\right\rangle
\label{Eq_27}
\end{equation}
where the subscripts \textit{u} and \textit{l} corresponds to the upper and lower waveguides, respectively. Thus both the signal and idler photons would  exit either from the upper waveguide (fundamental mode) or from the lower waveguide (first excited mode), resulting in path entanglement. 
If the SPDC process is chosen to lead to a degenerate pair of photons, then the output from the device would be a NOON state with \textit{N}=2. By modifying the waveguide parameters, the above scheme can be extended to generate  different combinations of  modal, polarization, path entangled state giving rise to a hyper-entangled state. Also, it can be used to generate a multipath entangled state by appropriately engineering the phase matching grating in the coupler region and enabling  multiple SPDC processes simultaneously. This would lead to the generation of higher order modes giving rise to a multi-mode, two photon entangled state. The output Y coupler can then be fabricated with more than two output ports, correspondingly giving rise to a multi-path, two photon entangled state.
\begin {figure}
\includegraphics[height=5.8cm]{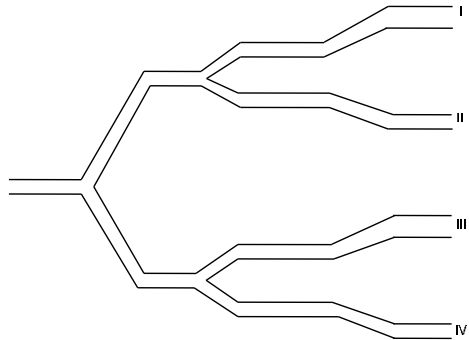}
\caption{Waveguide geometry for generating four path entangled state.}
\label{Fig_fourpath}
\end{figure}

   The multipath entangled state can also be achieved by combining more of these above geometries and designing the structure appropriately. As an example, a four path-two photon entangled state can be generated by designing the structure as shown in Fig.~\ref{Fig_fourpath}. Two of the waveguides of Fig.~\ref{Fig_wavegeom} are combined using a symmetric Y-coupler at the input and the signal-idler photon pair can exit from either of the four available paths with equal probability, giving rise to a maximally multi(four, in this case)-path entangled state. The potential applications for such multipath entangled states have already been discussed in Ref.~\cite{Rossi}.
\section{Conclusion}
\label{conclusion}
In this paper, we have addressed the issue of generation of  mode entangled photon pairs using optical waveguide geometry which has the inherent advantage of being compact (making the photons less susceptible to decoherence) and having enhanced optical non linearities in comparison to bulk crystal  or optical fibers. The proposed scheme is very efficient as the pump, signal and idler are taken to be all extraordinarily polarized, making use of the highest non linear coefficient $d_{33}$ of lithium niobate, and due to our choice of the geometry and the modes, the overlap integrals determining the efficiency of the two processes are also significant. We have shown that a maximally entangled state is achievable for a wide range of core separation of the coupler, with the additional advantage that one of the entangled photons is emitted at the typical wavelength of 1550 nm and can be used as a flying qubit. This mode entangled state can further be converted to a path entangled state which should  have varied applications in quantum lithography and quantum information processing. The proposal when extended to degenerate SPDC processes would lead to the production of a NOON state with \textit{N} = 2 photons. By appropriately choosing the waveguide parameters and also designing the integrated structure, the idea is applicable to produce a hyperentangled state and multi-path two photon entangled state which may have future applications in quantum computation protocols. 

\section*{Acknowledgement}
The work of JL was supported by University Grants Commission, New Delhi, India.

\end{document}